# Seismic reliability assessment of classical columnssubjected to near-fault ground motions

Ioannis N. Psycharis[1*], Michalis Fragiadakis[2], Ioannis Stefanou[3]

[1] *School of Civil Engineering, National Technical University of Athens, Greece*
[2] *Department of Civil and Environmental Engineering, University of Cyprus*
[3] *UR Navier-MSA, Ecole des Ponts ParisTech, Université Paris-Est, Marne-la-Vallée Cedex 2, France*

SUMMARY

A methodology for the performance-based seismic risk assessment of classical columns is presented. Despite their apparent instability, classical columns are, in general, earthquake resistant, as proven from the fact that many classical monuments have survived many strong earthquakes over the centuries. Nevertheless, the quantitative assessment of their reliability and the understanding of their dynamic behaviour are not easy, due to the fundamental non-linear character and the sensitivity of their response. In this paper, a seismic risk assessment is performed for a multidrum column using Monte Carlo simulation with synthetic ground motions. The ground motions adopted contain a high and a low frequency component, combining the stochastic method and a simple analytical pulse model to simulate the directivity pulse contained in near source ground motions. The deterministic model for the numerical analysis of the system is three dimensional and is based on the Discrete Element Method (3D DEM). Fragility curves are produced conditional on magnitude and distance from the fault and also on scalar intensity measures for two engineering demand parameters (*EDP*s), one concerning the intensity of the response during the ground shaking and the other the residual deformation of the column. Three performance levels are assigned to each *EDP*. Fragility analysis demonstrated some of the salient features of these spinal systems under near-fault seismic excitations, as for example their decreased vulnerability for very strong earthquakes of magnitude 7 or larger. The analysis provides useful results regarding the seismic reliability of classical monuments and decision making during restoration process.

## INTRODUCTION

Classical monuments are made of structural elements (drums in the case of columns), which lie one on top of the other without mortar. During a strong earthquake, the columns respond with intense rocking and, depending on the magnitude of the induced accelerations, sliding of the drums. In rare cases, steel connections (dowels) are provided at the joints, which restrict, up to their yielding, sliding but do not affect, in general, rocking.

[*] Corresponding author. Laboratory for Earthquake Engineering, National Technical University of Athens, HeroonPolytechneiou 9, Zografos 15780, Greece, Tel.: (+30) 2107721154, Fax: (+30) 2107721182, Email: ipsych@central.ntua.gr

Several investigators have examined the seismic response of classical monuments and, in general, of stacks of rigid bodies analytically, numerically or experimentally, mostly using two-dimensional models (e.g. [1] – [6] among others) and lesser using three-dimensional ones (e.g. [7] – [12]). Thesestudies have shown that the response is non-linear and sensitive even to small changes of the parameters. These characteristics are evident even to the simplest case of a rocking rigid block (Housner [13]).

Previous analyses of the seismic response of classical columns have shown that these structures, despite their apparent instability to horizontal loads, are, in general, earthquake resistant (Psycharis *et al*. [5]), which is also proven from the fact that many classical monuments built in seismic prone areas have survived for almost 2500 years. However, many others have collapsed.

In general, the vulnerability of ancient monuments to earthquakes depends on two main parameters (Psycharis, *et al*. [5]): the size of the structure and the predominant period of the ground motion. Concerning the size, larger columns are more stable than smaller ones with the same aspect ratio of dimensions. Concerning the period of the excitation, it affects significantly the response and the possibility of collapse, with low-frequency earthquakes being much more dangerous than high-frequency ones. In this sense, near field ground motions, which contain long-period directivity pulses, might bring these structures to collapse.

The assessment of the seismic reliability of a monument is a prerequisite for the correct decision making during a restoration process. The seismic vulnerability of the column, not only in what concerns the collapse risk, but also the magnitude of the expected maximum and residual displacements of the drums, is vital information that can help the authorities decide the necessary interventions. This assessment is not straightforward, not only because fully accurate analyses for the near-collapse state are practically impossible due to the sensitivity of the response to small changes in the geometry and the difficulty in modelling accurately the existing imperfections, but also because the results depend highly on the ground motions characteristics.

It is evident therefore that the assessment of the seismic reliability of monuments will improve our understanding of how these systems have survived over the centuries and will also help to prioritize future interventions. This task is not trivial and requires expanding our understanding of performance-based design concepts for the capacity assessment of ancient monuments.

In this paper, a risk assessment is performed for the case study of a column of the Parthenon Pronaos in Athens, Greece. To this end, we present a vulnerability assessment approach that accounts for the record-to-record variability. Record-to-record variability is also termed "aleatory uncertainty", or "randomness", and is responsible for significant variability in the seismic response. Advanced modelling and numerical analysis tools are combined with performance-based earthquake engineering concepts. The performance-based concept is expanded to classical monuments adopting appropriate performance levels and demand parameters to develop a decision-support system that will take into consideration engineering parameters helping the authorities on deciding upon the interventions required.

It is noticed that the Parthenon column is used only as an example for the application of the proposed methodology and that the analysis that is presented does not aim to evaluate the vulnerability of Parthenon. The specific column was chosen as a typical example, as it represents a medium-size column of common slenderness. For this reason, damage was not introduced in the model and the analysis was not restricted only to earthquakes that can occur in Athens. Therefore, many of the conclusions drawn can be generalized. It must be emphasized, however, that the results presented herein cannot be applied quantitatively in all cases. A proper vulnerability analysis is required on a case-by-case basis, taking under consideration the geometry of the column under consideration, the existing damage and the seismotectonic environment of thesite.

## NUMERICAL MODELLING OF MULTIDRUM COLUMNS

During a seismic event, the response of a multidrum column is dominated by the "spinal" form of the construction and is governed by the sliding, the rocking and the wobbling of the individual, practically rigid, stone drums, which translate and rotate independently or in groups (Figure 1). There are many 'modes' in which the system can vibrate, with different joints being opened in each mode, and the column continuously moves from one oscillation 'mode' to another. The term 'mode' is used here to denote different patterns of the response and does not refer to the eigenmodes of the system, since spinal structures do not possess natural modes in the typical sense.

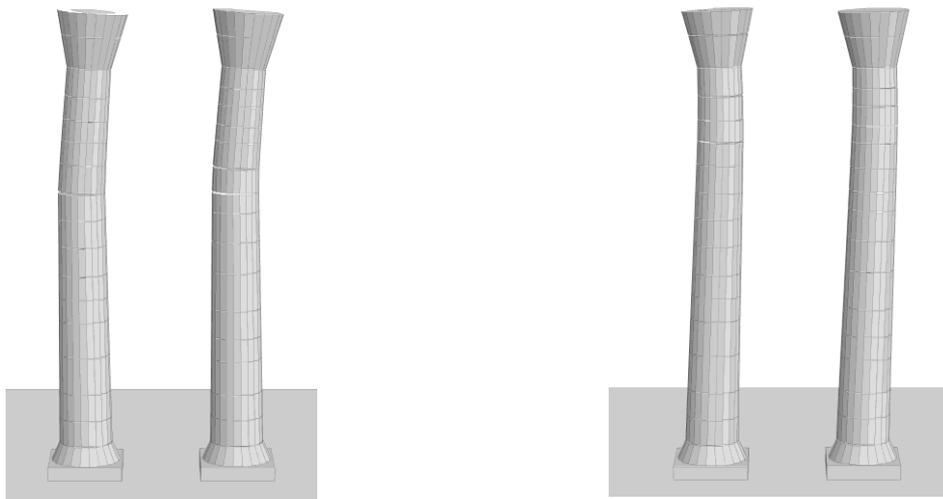

Figure 1.Response of two columns of Olympieion of Athens at two different time instances during intense ground shaking. The geometry of the two columns is slightly different (the left has 14 drums and the right 15) leading to different 'modes' of vibration (numerical results obtained with 3DEC software).

The underlying mathematical problem is strongly non-linear and consequently the modelling of the dynamic behaviour of multidrum columns is quite complex. Even in the case of systems with a single-degree-of-freedom in the two dimensional space, i.e. a monolithic rocking block, the analytical and the numerical analysis is not trivial (Housner [13]) and differs from the approaches followed in modern structural analysis. The dynamic response becomes even more complex in three dimensions, where realistic models have to account for several non-linearities related to the three dimensional motion of each drum and the energy dissipation at the joints. For a more extensive discussion on the dynamic behaviour of such spinal systems we refer to Psycharis [3], Mouzakis *et al.* [8], Dasiou *et al.* [10], Stefanou*et al*. [14] among others.

Herein, we used the Discrete (or Distinct) Element Method (DEM) for the numerical modelling of the seismic response of multidrum systems. DEM may not be the only choice for the discrete system at hand, but it forms an efficient and validated manner for the study of the dynamic behaviour of masonry columns in classical monuments. The Molecular Dynamics (smooth-contact) approach was followed here (Cundall & Strack [15]) and the three dimensional DEM code 3DEC (Itasca [16]) was used. This software code provides the means to apply the conceptual model of a masonry structure as a system of blocks which may be considered either rigid, or deformable. In the present study only rigid blocks were used, as this was found to be a sufficient approximation and capable to reduce substantially the computing time. The system deformation is concentrated at the joints (soft-contacts), where frictional sliding and/or complete separation may take place (dislocations and/or disclinations between blocks). As discussed in more detail by Papantonopoulos *et al.*[7], the discrete element method employs an explicit algorithm for the solution of the equations of motion, taking into account large displacements and rotations. The efficiency of the method and particularly of 3DEC to capture the seismic response of classical structures has been already examined by juxtaposing the numerical results with experimental data (Papantonopoulos, *et al.* [7]; Dasiou, *et al.* [11]).

The geometry of the column considered in the present study was inspired by the columns of the Parthenon Pronaos on the Acropolis Hill in Athens. The column has a total height of 10.08 m, being composed of a shaft of 9.38 m and a capital. The real column has 20 flutes; however, the shaft in the numerical model was represented in an approximate manner by a pyramidal segment made of blocks of polygonal 10-sided cross section with diameters ranging from 1.65 m at the base to 1.28 m at the top. The shaft was divided into 12 drums of different height according to actual measurements of the columns of the Pronaos (Figure 2).

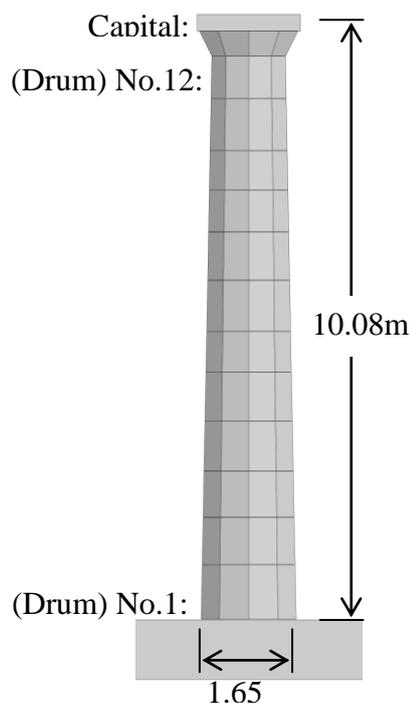

Figure 2. The multidrum column considered in the analyses.

A quite important factor for the numerical analysis is the selection of the appropriate constitutive laws that govern the mechanical behaviour of the joints. In the present paper we made use of a Coulomb-type failure criterion. In Table I we list the friction angle, the cohesion, the ultimate tensile strength and the stiffness of the joints. Notice that the stiffness might affect considerably the results of the analysis. A parametric investigationperformed by Toumbakari & Psycharis [17] showed that stiff joints might lead to larger permanent dislocations of all drums for strong ground motions compared with joints of soft stiffness. The values presented in Table I correspond to marble columns and were calibrated against shaking table experiments on the earthquake response of free-standing columns(Papantonopoulos, *et al.* [7]); with these values, good agreement was achieved concerning both the maximum top displacement and the residual displacements of the drums. It must be pointed out, however, that different values should be assigned to the stiffness parameters for material other than marble of good quality. One way to calculate the appropriate value for a specific column is by calibrating the stiffness against ambient vibration measurements (e.g. see [18]).

No artificial (numerical) damping was introduced to the system. According to the results of a previous investigation (Papantonopoulos, et al. [7]), damping is set to zero only during the intense rocking response, while non-zero damping is considered after that period in order to dissipate the free vibrations and make possible to determine the permanent deformation. However, a previous investigation (Toumbakari and Psycharis [17]) showed that, in general, the value of the damping that is used after the strong motion and the time at which it is introduced do not affect significantly the response of the column and the residual displacements. Based on this conclusion, zero damping was considered in the present analysis for the whole time history of the response in order to minimize the runtimes, as damping generally decreases the time step increasing the runtime. Since the free rocking oscillations after the end of the strong ground motion were not dissipated, the residual deformation of the column was calculated from the average displacements of the drums in the last two seconds of the response. The validity of this approach was verified for five earthquake records, for which the residual displacements of the drums were calculated: (i) with the above-mentioned procedure; and (ii) after the introduction of significant mass-proportional damping after the end of the strong ground motion. In the latter case, instead of stiffness-proportional, mass-proportional damping was preferred, since the introduction of stiffness-proportional damping was leading to extremely small integration steps, thus making the analysis practically impossible. The results are presented in Table II and show an error ranging from 2% to 18%, which is considered acceptable taking under consideration the sensitivity of the response.

Table I. Constitutive parameters for the Coulomb elastoplastic model considered for the mechanical behaviour of the joints.

| | |
|---|---|
| Normal Stiffness | 1 GPa/m |
| Shear Stiffness | 1 GPa/m |
| Friction Angle | 37° |
| Cohesion | 0 MPa |
| Tensile strength | 0 MPa |

Table II. Comparison of the maximum residual displacement of the drums without considering any damping and introducing mass-proportional damping after the end of the strong ground shaking.

| Earthquake | Normalized maximum residual drum dislocation, $u_d$ [*] | | |
|---|---|---|---|
| | Mass damping | No damping | Error (%) |
| GAZLI | $1.226 \times 10^{-2}$ | $1.450 \times 10^{-2}$ | 18.2 |
| SAN SALVADOR | $0.906 \times 10^{-2}$ | $0.992 \times 10^{-2}$ | 9.5 |
| ERZICAN | $0.651 \times 10^{-2}$ | $0.614 \times 10^{-2}$ | −5.6 |
| NORTHRIDGE (JFA) | $0.770 \times 10^{-2}$ | $0.645 \times 10^{-2}$ | −16.3 |
| CHICHI | $0.868 \times 10^{-2}$ | $0.886 \times 10^{-2}$ | 2.1 |

[*] Normalization with respect to the drum diameter: $u_d = \max(\mathrm{res}\, u_i)/D_i$.

No connections were considered between the drums, as the only connectors present in the original structure are wooden dowels, the so-called 'empolia', which were used to centre the drums during the erection of the column and not to provide a shear resistant mechanism. The shear strength of the wooden dowels is small and has only marginal effect to the response of the column (Konstantinidis and Makris [6]); for this reason, the wooden dowels were not considered in the numerical model.

## FRAGILITY ASSESSMENT

Fragility (or vulnerability) curves are a valuable tool for the seismic risk assessment of a system. Fragility analysis was initially developed for the reliability analysis of nuclear plants in an effort to separate the structural analysis part from the hazard analysis performed by engineering seismologists. Vulnerability analysis requires the calculation of the probabilities that a number of monotonically increasing limit-states are exceeded. Therefore, the seismic fragility $F_R$ is defined as the limit-state probability conditioned on seismic intensity. The seismic intensity can be expressed in terms of magnitude $M_w$ and distance $R$, resulting to a surface $F_R(M_w,R)$. Therefore, the fragility of a system is the probability that an engineering demand parameter (*EDP*) exceeds a threshold value *edp* and is defined as:

$$F_R(M_w, R) = P(EDP \geq edp | M_w, R) \qquad (1)$$

Eq. (1) provides a single-point of a limit-state fragility surface, while engineering demand parameters (*EDPs*) are quantities that characterize the system response, e.g., permanent or maximum deformation, drum dislocation. To calculate $F_R$ we performed Monte Carlo Simulation (MCS) using Latin Hypercube Sampling (LHS) for a range of magnitude and distance ($M_w$, $R$) scenarios. For this purpose, a large number of nonlinear response history analyses for every $M_w$–$R$ pair is needed, especially when small probabilities are sought. Therefore, suites of records that correspond to the same $M_w$ and $R$ value must be compiled. Since it is very difficult to come up with such suites of natural ground motion records, we produced synthetic ground motions following the procedure discussed in the following section.

Assuming that seismic data are lognormally distributed (Benjamin & Cornell [19]; Shome *et al.* [20]), $F_R(M_w,R)$ can be calculated analytically once the mean and the standard deviation of the logs of the *EDP* are calculated, which are denoted as $\mu_{\ln EDP}$ and $\beta_{\ln EDP}$, respectively. Once they are known they can be used to calculate $F_R$ using the normal distribution:

$$F_R = P(EDP \geq edp | M_w, R) = 1 - \Phi\left(\frac{\ln(edp) - \mu_{\ln EDP}}{\beta_{\ln EDP}}\right) \qquad (2)$$

where *edp* is the *EDP*'s threshold value that denotes that the limit-state examined is violated and $\Phi$ denotes the standard normal distribution. For example, if we are calculating the fragility surface that corresponds to normalized displacement of the column's capital $u_{\text{top}}$ (defined in the ensuing) larger than 0.3, $\ln(edp)$ would be equal to $\ln(0.3)$. Alternatively, a good approximation of Eq. (1) can be obtained by the ratio of successful simulations over the total number of simulations performed, thus bypassing the assumption of lognormality. For the case study examined in this paper, the two approaches give results that practically are close.

As the ground motion intensity increases, some records may result in collapse of the structure. When collapsed simulations exist, Eq. (2) is not accurate, since the *EDP* takes an infinite or a very large value that cannot be used to calculate $\mu_{\ln EDP}$ and $\beta_{\ln EDP}$. To handle such cases, Eq. (2) is modified by separating the data to collapsed and non-collapsed ones. The conditional probability of collapse is calculated as:

$$P(C|M_w,R) = \frac{\text{number of simulations collapsed}}{\text{total number of simulations}} \qquad (3)$$

If $\mu_{\ln EDP}$ and $\beta_{\ln EDP}$ are the mean and the dispersion of the non-collapsed data respectively, Eq. (2) is modified as follows:

$$P(EDP \geq edp|M_w,R) = P(C|M_w,R) + (1-P(C|M_w,R)) \cdot \left(1 - \Phi\left(\frac{\ln(edp) - \mu_{\ln EDP}}{\beta_{\ln EDP}}\right)\right) \qquad (4)$$

It is also customary to produce fragility curves using a single scalar intensity measure *IM*. Thus, instead of conditioning $F_R$ on magnitude and distance (Eq. (1)) we can use a scalar intensity measure *IM* resulting to a fragility curve $F_R(IM)$. Typical intensity measures are the peak ground acceleration (*PGA*), the peak ground velocity (*PGV*), the spectral acceleration (*SA*), the spectral velocity (*SV*), or any other variable that is consistent with the specification of seismic hazard. This option is often preferred, not only because 2D plots are easier to interpret than three-dimensional surfaces but, mainly, because this option is easier in terms of handling the ground motion records. In order to calculate conditional probabilities, usually the ground motions are scaled at the same *IM* value. Since record scaling is a thorny issue that may introduce biased response estimates, this option was not preferred.

Fragility curves can be alternatively produced with smart post-processing of the data. If the data, regardless of their $M_w$ and *R* value, are plotted in *EDP–IM* ordinates (Figure 3) the conditional probabilities can be calculated by dividing the *IM* axis into stripes, as shown on Figure 3. If $IM_m$ is the *IM* value of the stripe, the conditional probability $P(EDP \geq edp|IM_m)$ can be calculated according to Eq. (2) or (4) using only the data banded within the stripe. Thus, according to Figure 3, if the moving average $\mu_{\ln EDP}$ and the dispersion $\beta_{\ln EDP}$ are calculated using only the black dots, $P(EDP \geq edp|IM_m)$ can be approximately calculated using Eq. (4). Some readers may assume that the coupling between $M_w$–*R* and an *IM* can be easily obtained using a groundmotion prediction equation, also known as attenuation relationship. However, this should be avoided, since groundmotion prediction equations have significant scatter and they have not been derived to serve such purposes.

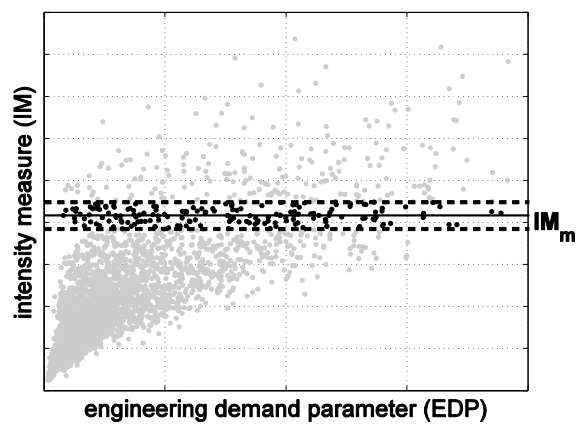

Figure 3. Post-processing to obtain fragility curves from scattered data.

GENERATION OF SYNTHETIC, HAZARD–CONSISTENT GROUND MOTIONS

The assessment of the seismic reliability of the column of Parthenon that is presented herein is based on synthetic ground motions, representative of near-field sites. The reason of using synthetic instead of natural ground motions, is the limited number of the latter for the range of pairs $M_w$–$R$ that are examined, especially for stiff soil conditions on which monuments are typically founded. The synthetic records were generated using the process that has been proposed by MavroeidisandPapageorgiou[21], which allows for the combination of independent models that describe the low-frequency (long period) component of the directivity pulse, with models that describe the high-frequency component of an acceleration timehistory. A successful application of this approach is given in Taflanidis*et al.*[22]. In the present paper, the generation of the high-frequency component was based on the stochastic (or engineering) approach discussed in detail in Boore[23]. Based on a given magnitude-distance scenario ($M_w$–$R$) and depending on a number of site characteristics, the stochastic approach produces synthetic ground motions.

It must be noted that, due to the high nonlinear nature of the rocking/wobbling response and the existence of a minimum value of the peak ground acceleration that is required for the initiation of rocking, the high frequency part of the records is necessary for the correct simulation of surrogate ground motions. Long-period directivity pulses alone, although they generally produce devastating effects to classical monuments (Psycharis, *et al.* [5]), might not be capable to produce intense shaking and collapse, as the maximum acceleration of pulses of long period is usually small and not strong enough to even initiate rocking.

Classical monuments were usually constructed on the Acropolis of ancient cities, i.e. on top of cliffs; thus, most of them are founded on stiff soil or rock, and only few of them on soft soil. For this reason, the effect of the soil on the characteristics of the exciting ground motion was not considered in the present analysis. It is noted, though, that, although the directivity pulse contained in near-fault records is not generally affected by the soil conditions, soft soil can significantly alter the frequency content of the ground motion and, consequently, affect the response of classical columns. This effect, however, is beyond the scope of this paper.

*Low frequency pulse*

For the long-period component of the synthetic ground motions we applied the pulse model of MavroeidisandPapageorgiou[21]. This wavelet has been calibrated using actual near-field ground motions from all-over the world. The velocity pulse is given by the expression:

$$V(t)=0.5A_p\left[1+\cos\left(\frac{2\pi f_p}{\gamma_p}(t-t_0)\right)\right]\cos\left[2\pi f_p(t-t_0)+v_p\right], \quad t\in\left[t_0-\frac{\gamma_p}{2f_p}, t_0+\frac{\gamma_p}{2f_p}\right] \quad (5)$$

where $A_p$, $f_p$, $v_p$, $\gamma_p$ and $t_0$ describe the amplitude of the envelope of the pulse, the prevailing frequency, the phase angle, the oscillatory character (i.e., number of half cycles) and the time shift to specify the epoch of the envelope's peak, respectively. All parameters of Eq. (5) have a clear and unambiguous meaning. For every magnitude–distance scenario ($M_w$–$R$), the velocity amplitude of the directivity pulse ($V_p$) and the frequency $f_p$ were obtained using the expressions produced by Rupakhety*et al.* [24]. Specifically, the meanvalue of $V_p$ was obtained by:

$$\log(V_p)=-5.17+1.98\cdot M_w-0.14\cdot M_w^2-0.10\cdot\log(R^2+0.562) \quad (6)$$

where $M_w$ cannot exceed $M_{sat}$, which is considered equal to 7.0. Thus, for magnitude values above $M_{sat}$, we set $M_w = M_{sat}$ to obtain $V_p$ using Eq. (6). Similarly, the mean pulse frequency $f_p$ is:

$$\log(1/f_p) = -2.87 + 0.47 \cdot M_w \quad (7)$$

Note that equations (6) and (7) use base 10 logarithms. Also, $V_p$ is not in general equal to the envelope amplitude $A_p$, but one can be calculated from the other if the phase angle $v_p$ is known.

We randomly constructed low-frequency pulse-like ground motions using Eq.(5) and giving random values to $V_p$, $f_p$, $v_p$ and $\gamma_p$. Sets of pulse-like ground motions were obtained for every $M_w$–$R$ combination using Latin Hypercube Sampling. We assumed that the logarithms of $V_p$ and $f_p$ follow the normal distribution with standard deviation equal to 0.16 and 0.18, respectively (Rupakhety, *et al.* [24]). The phase angle $v_p$ was randomly chosen in the [−π/2, π/2] range. Moreover, being consistent with the data of Mavroeidis & Papageorgiou [21], the number of half cycles $\gamma_p$ was assumed to follow a normal distribution with mean and standard deviation equal to 1.8 and 0.4, respectively. The distribution of $\gamma_p$ was left-truncated to one, while $V_p$ and $f_p$ were also left-truncated to zero, ensuring that no negative values were sampled. Figure 4 shows the histogram of the four random parameters used for creating pulses for the $M_w = 7$ and $R = 5$ km case.

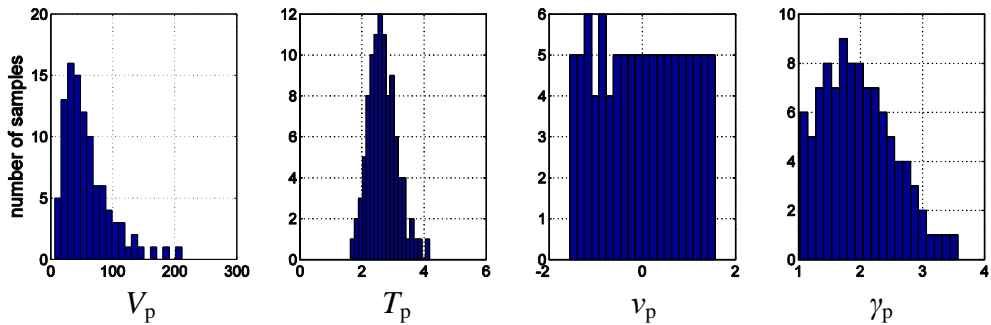

Figure 4. Histogram of the random parameters that describe the low-frequency plot ($M_w = 7$ and $R = 5$ km).

*High frequency component – The stochastic approach*

The stochastic approach was selected for modelling the high-frequency component of the ground motions. The stochastic method is discussed in detail in Boore[23] and is based on the ground motion radiation spectrum $Y(M_w,R,f)$, which is the product of quantities that consider the effect of source, path, site and instrument (or type) of motion. By separating the spectrum to its contributing components, the models based on the stochastic method can be easily modified to account for different problem characteristics. The shape and the duration of the ground motions depend on an envelope function $w(M_w,R,t)$.

The steps followed to generate the high frequency component are briefly summarized as follows. First generate white noise (Gaussian or uniform) for a duration given by the duration of the motion as predicted by an appropriate ground motion prediction equation. The noise is then windowed and transformed into the frequency domain using the envelope function $w(M_w,R,t)$. The spectrum is normalized by the square root of the mean square amplitude spectrum and multiplied by the ground motion spectrum $Y(M_w,R,f)$. The resulting spectrum is transformed back to the time domain. The $Y(M_w,R,f)$ spectrum and the model parameters adopted in our study are these of Atkinson & Silva [25]. All simulations have been performed using the SMSIM program, freely available from http://www.daveboore.com.

*Combined synthetic strong ground motions*

Synthetic ground motion records were constructed for magnitudes $M_w$ in the range 5.5 to 7.5 with a step of 0.5 (five values of $M_w$) and distances from the fault $R$ in the range 5 to 20 km with a step of 2.5 km (seven values of $R$). In total, 35 pairs of $M_w$–$R$ were considered. For each $M_w$–$R$ scenario, 100 Monte Carlo Simulations (MCS) were performed for a random sample of $V_p, f_p, v_p, \gamma_p$ using Latin Hypercube Sampling to produce the low-frequency pulse, while the high-frequency component was produced using the stochastic method, producing thus 100 random ground motions compatible with the $M_w$–$R$ scenario considered.

The procedure we used to combine the low and high frequency components is shown schematically in Figure 5. The steps are as follows:
1. Apply the stochastic method to generate an acceleration time history to use as the high-frequency component for a given moment magnitude $M_w$ and distance $R$ scenario.
2. For the $M_w$–$R$ scenario considered, sample $V_p, f_p, v_p, \gamma_p$ and obtain the low-frequency directivity pulse using Eq.(5). Shift the pulse so that its maximum velocity coincides in time with the maximum of the velocity time history of the high-frequency record of Step 1.
3. Calculate the Fourier transform of both high- and low-frequency components. Calculate also the phase angle of the high-frequency component.
4. Subtract the Fourier amplitude of the pulse from that of the high-frequency component of the ground motion.
5. Construct a synthetic acceleration time history so that its Fourier amplitude is that of Step 4 and its phase angle is that of the high-frequency record calculated in Step 3.
6. The final synthetic record is obtained by adding the pulse time history and the time history of Step 5.

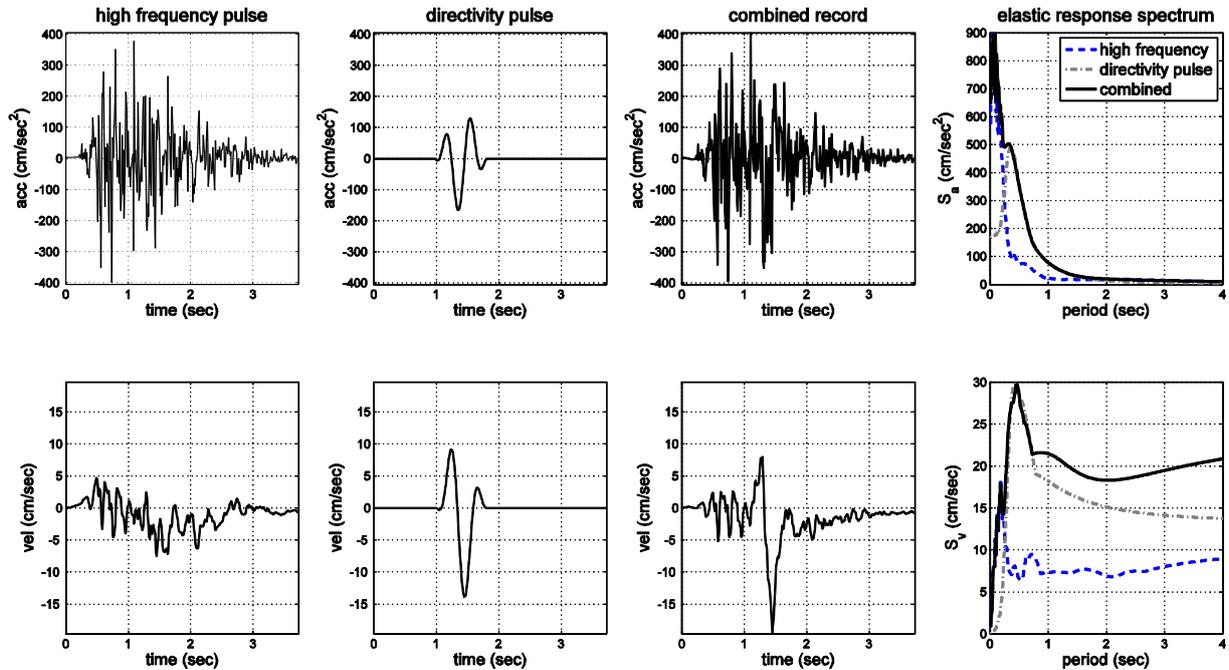

Figure 5. Generation of synthetic ground motion records. Upper row shows acceleration and the bottom row the velocity timehistories and response spectra.

The last column of Figure 5 shows the corresponding acceleration and velocity response spectra. The velocity spectrum (bottom right figure) shows the impact of the directivity pulse for the period range around $1/f_p$. Moreover, looking at the third column, the effect of the pulse is clearly visible in the combined velocity timehistory but difficult to identify when looking at the acceleration timehistory.

## PERFORMANCE-BASED RELIABILITY ASSESSMENT OF CLASSICAL MONUMENTS

Performance-Based Earthquake Engineering (PBEE) and seismic risk assessment combine computational tools and reliabilityassessment procedures to obtain the system fragility for a wide range of limit states. The seismic risk assessment requires the calculation of the failure probabilities of a pre-set number of performance objectives. According to PBEE, the acceptable level of damage sustained by a structural system depends on the level of ground shaking and its significance. For example, under a frequent earthquake a building should be able to tolerate minor, non-structural, damage, but a critical facility (e.g. a bridge or a hospital) should remain intact and fully operable. Thus, the target in risk assessment is to obtain the probabilities of violating the stated performance levels, ranging from little or no damage for frequent earthquakes to severe damage for rare events. Today, these concepts are well understood among earthquake engineers, but when classical monuments are considered the performance-based criteria may differ considerably. For example, to retrofit an ancient column one has to decide what is the 'acceptable level' of damage for a given intensity level. The approach for making such decisions is not straightforward. A consensus among various experts in archaeology and monument preservation is necessary, while a number of non-engineering decisions have to be taken.

In order to assess the risk of a monument, the performance levels of interest and the corresponding levels of capacity of the monument need first to be decided. Demand and capacity should be measured with appropriate parameters (e.g. stresses, strains, displacements) at critical locations, in accordance to the different damage (or failure) modes of the structure. Subsequently, this information has to be translated into one or a combination of engineering demand parameters (*EDP*s), e.g., permanent or maximum column deformation, drum dislocation, foundation rotation or maximum axial and shear stresses. For the *EDP*s chosen, appropriate threshold values that define the various performance objectives e.g. light damage, collapse prevention, etc. need to be established. Since such threshold valuesare not always directly related to visible damage, the *EDP*s should be related to damage that is expressed in simpler terms, e.g., crack width, crack density or exfoliation surface area. In all, this is a challenging, multi-disciplinary task that requires experimental verification, expert opinion and rigorous formulation.

In the investigation presented here, two engineering demand parameters (*EDP*s) are introduced for the assessment of the vulnerability of classical columns: (a) the maximum displacement at the capital normalized by the base diameter (lower diameter of drum No. 1, see Fig. 2); and (b) the relative residual dislocation of adjacent drums normalized by the diameter of the corresponding drums at their interface. The first *EDP* is the maximum of the normalized displacement of the capital (top displacement) over the whole timehistory and is denoted as $u_{top}$, i.e. $u_{top} = \max[u(top)]/D_{base}$. This is a parameter that provides a measure of how much a column has been deformed during the ground shaking and also shows how close to collapse the column was brought during the earthquake. Note that the top displacement usually corresponds to the maximum displacement among all drums. The second *EDP* is the residual relative drum dislocations at the end of the seismic motion normalised by the drum diameter at the corresponding joints and is denoted as $u_d$, i.e. $u_d = \max(\text{res}u_i)/D_i$. This parameter provides a measure of how much the geometry of the column has been altered after the earthquake increasing thus the vulnerability of the column to future events.

The *EDP*s proposed have a clear physical meaning and allow to easily identify various damage states and setting empirical performance objectives. For example a $u_{top}$ value equal to 0.3 indicates that the maximum displacement was 1/3 of the bottom drum diameter and thus there was no danger of collapse, while values of $u_{top}$ larger than unity imply intense shaking and large deformations of the column, which, however, do not necessarily lead to collapse. It is not easy to assign a specific value of $u_{top}$ that corresponds to collapse, as collapse depends on the 'mode' of deformation, which in turn depends on the ground motion characteristics. For example, for a cylindrical column that responds as a monolithic block with a pivot point at the corner of its base (Figure 6a), collapse is probable to occur for $u_{top} > 1$, as the weight of the column turns to an overturning force from a restoring one when $u_{top}$ becomes larger than unity. But, if the same column responds as a multidrum one with rocking at all joints (Figure 6b), a larger value of $u_{top}$ can be attained without threatening the overall stability. In fact, the top displacement can be larger than the base diameter without collapse, as long as the weight of each part of the column above an opening joint gives a restoring moment about the pole of rotation of the specific part. In the numerical analyses presented here, the maximum value of $u_{top}$ that was attained without collapse was about 1.15.

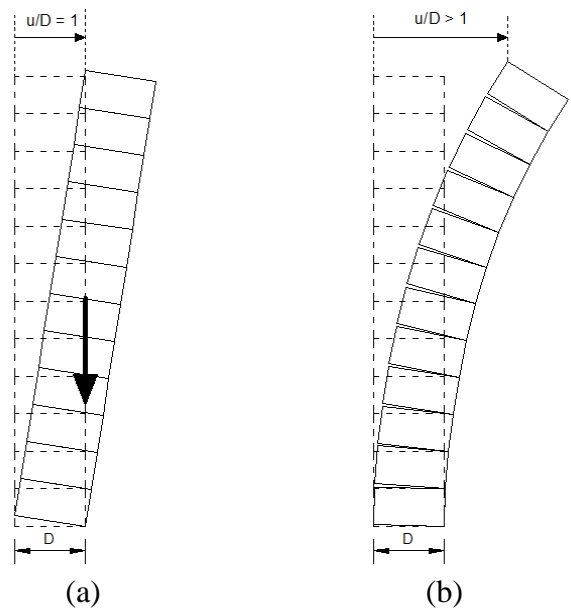

Figure 6. Top displacement for two extreme modes of rocking: (a) as a monolithic block; (b) with opening of all joints (displacements are shown exaggerated).

Based on the above defined *EDP*s, the performance criteria of Tables III and IV have been adopted. For $u_{top}$, three performance levels were selected (Table III), similarly to the ones that are typically assigned to modern structures. The first level (*damage limitation*) corresponds to weak shaking of the column with very small or no rocking. At this level of shaking, no damage, nor any severe residual deformations are expected. The second level (*significant damage*) corresponds to intense shaking with significant rocking and evident residual deformation of the column after the earthquake; however, the column is not brought close to collapse. The third performance level (*near collapse*) corresponds to very intense shaking with significant rocking and probably sliding of the drums. The column does not collapse at this level, as $u_{top} < 1$, but it is brought close to collapse. In most cases, collapse occurred when this performance level was exceeded. The values of $u_{\text{top}}$ that are assigned at every performance level are based on the average assumed risk of collapse.

TableIII. Proposed performance criteria concerning the risk of collapse.

| $u_{\text{top}}$ | Performance level | Description |
| --- | --- | --- |
| 0.15 | Damage limitation | No danger for the column. No permanent drum dislocations expected. |
| 0.35 | Significant damage | Large opening of the joints with probable damage due to impacts and considerable residual dislocation of the drums. No serious danger of collapse. |
| 1.00 | Near collapse | Very large opening of the joints, close to partial or total collapse. |

TableIV. Proposed performance criteria concerning permanent deformation (residual drum dislocations).

| $u_{\text{d}}$ | Performance level | Description |
| --- | --- | --- |
| 0.005 | Limited deformation | Insignificant residual drum dislocations without serious effect to future earthquakes. |
| 0.01 | Light deformation | Small drum dislocations with probable unfavourable effect to future earthquakes. |
| 0.02 | Significant deformation | Large residual drum dislocations that increase significantly the danger of collapse during future earthquakes. |

Three performance levels were also assigned to the normalised residual drum dislocation, $u_{\text{d}}$ (Table IV). This *EDP* is not directly related to how close to collapse the column was brought during the earthquake, since residual displacements are caused by wobbling and sliding and are not, practically, affected by the amplitude of the rocking. However, their importance to the response of the column to future earthquakes is significant, as previous damage/dislocation has generally an unfavourable effect to the seismic response to future events (Psycharis [26]).

The first performance level (*limited deformation*) concerns very small residual deformation which is not expected to affect considerably the response of the column to future earthquakes. The second level (*light deformation*) corresponds to considerable drum dislocations that might affect the dynamic behaviour of the column to forthcoming earthquakes, increasing its vulnerability. The third performance level (*significant deformation*) refers to large permanent displacements at the joints that increase considerably the danger of collapse to future strong seismic motions. It must be noted, however, that the threshold values assigned to $u_{\text{d}}$ are not obvious, as the effect of pre-existing damage to the dynamic response of the column varies significantly according to the column properties and the characteristics of the ground motion. The values proposed are based on engineering judgment taking into consideration the size of drum dislocations that have been observed in monuments and also the experience of the authors from previous numerical analyses and experimental tests. Moreover, after quickly examining the results of this study it was observed that the first limit case was exceeded by most of the records examined, while the third case was exceeded only by a few ground motions.

The comparison of the two *EDP*s using all ground motions considered, apart from those that caused collapse, is shown in Figure 7. Although there is a clear trend showing that, generally, strong ground motions lead to large top displacements $u_{top}$ during the strong shaking and also produce large permanent deformation $u_d$ of the column, there is significant scattering of the results indicating that intense rocking does not necessarily imply large residual dislocations of the drums and also that large drum dislocations can occur for relatively weak shaking of the column. This was also observed during shaking table experiments (Mouzakis *et al*. [8]) where cases of intense rocking with very small residual drum displacements have been identified.

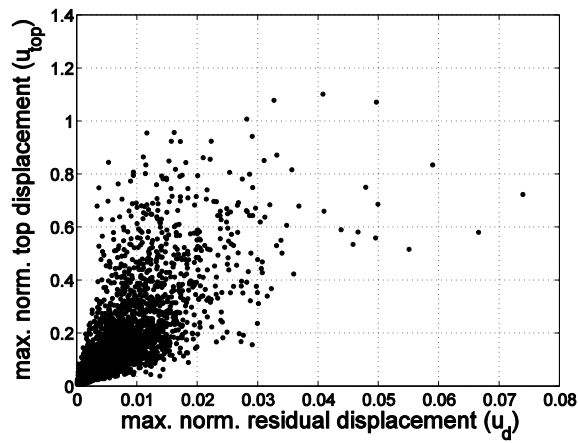

Figure 7. Comparison of $u_d$ versus $u_{top}$ for the ground motions which did not cause collapse.

## FRAGILITY CURVES

The proposed fragility assessment methodology was applied to the classical column of Figure 2. The response of the column was calculated for 35 $M_w$–$R$ scenarios. For every $M_w$–$R$ scenario 100 Monte Carlo Simulations (MCS) were performed, thus resulting to 3500 simulations in total.

Figure 8a shows the mean $u_{top}$ displacements of the column and Figure 8b the corresponding $u_d$ displacements. The surface plots of Figure 8a and 8b refer to non-collapsed simulations, while the collapse probabilities as function of magnitude and distance are shown in Figure 8c. Collapse is considered independently of whether it is local (collapse of a few top drums) or total (collapse of the whole column). As expected, the number of collapses is larger for smaller fault distances and larger magnitudes. For example, for $M_w = 7.5$ and $R = 5$ km 40% of the simulations caused collapse, while practically zero collapses occurred for magnitudes less than 6.5.

An unexpected behaviour is depicted in Figure 8. Concerning the mean top displacement during the seismic motion, Figure 8ashows that for small distances from the fault, up to approximately 7.5 km, the mean value of $u_{top}$ increases monotonically with the magnitude as expected. However, for larger fault distances, the maximum $u_{top}$ occurs for magnitude $M_w = 6.5$, while for larger magnitudes the top displacement decreases. For example, for $R = 20$ km, the mean value of $u_{top}$ is approximately 0.4 for $M_w = 6.5$, while the corresponding value for $M_w = 7.5$ is about 0.2, i.e. it is reduced to one half. This counter-intuitive response is attributed to the saturation of the *PGV* for earthquakes with magnitude larger than $M_{sat} = 7.0$ (Rupakhety *et al.* [24], see Eq. (6)) while the period of the pulse is increasing exponentially with the magnitude. As a result, the directivity pulse has small acceleration amplitude for large magnitudes, which is not capable to produce intense rocking. This is shown in Fig. 9, where the mean value of the the velocity amplitude, $V_p$, and acceleration amplitude, $A_p$, of the directivity pulse according to Eqs (6) and (7) are plotted versus $M_w$ and $R$.

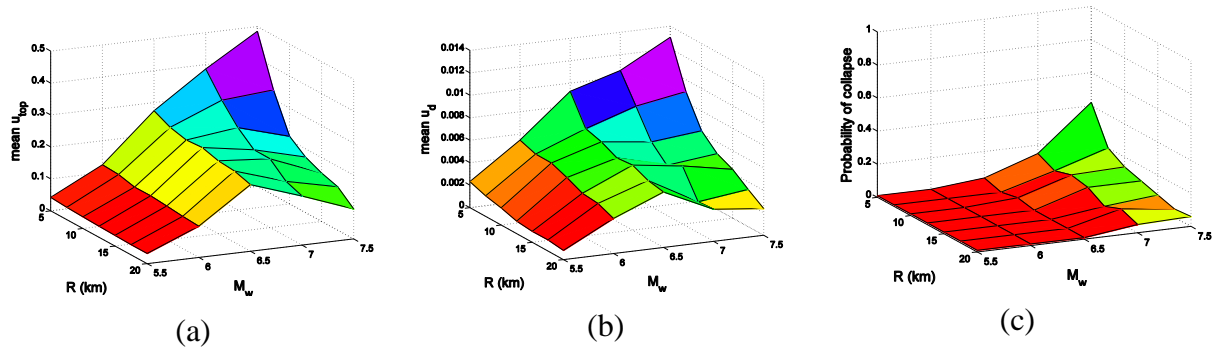

(a)          (b)          (c)

Figure 8. Mean values of the adopted *EDP*s for the classical column considered: (a) maximum normalised top displacements, $u_{top}$; (b) normalised residual deformations, $u_d$; and (c) Collapse probabilities for the multidrum column considered.

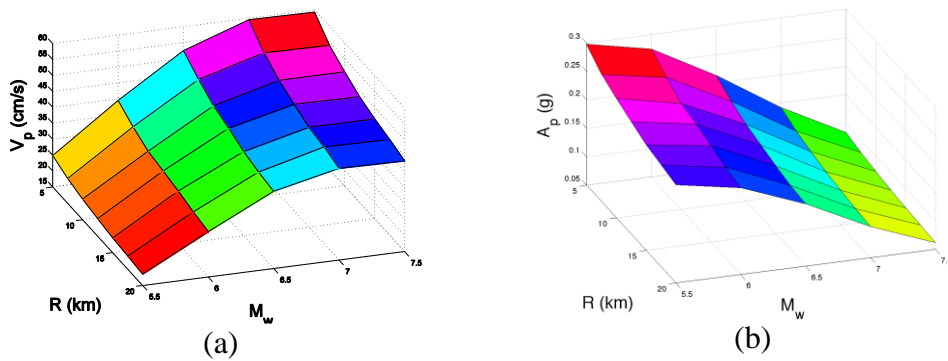

(a)          (b)

Figure 9. Variation of the mean value of: (a) the velocity amplitude and (b) the acceleration amplitude of the directivity pulse, according to Eqs (6) and (7), with the magnitude, $M_w$, and the distance, $R$, respectively.

Similar, and probably more pronounced, is the behaviour concerning the permanent drum dislocations $u_d$ shown in Figure 8b. Again, $u_d$ increases monotonically with the magnitude for small values of $R$ only, less than 10 km. For larger distances, $u_d$ increases with $M_w$ up to magnitudes equal to 6.5, when it attains its maximum value. For larger magnitudes smaller permanent deformation of the column occurs.

To verify the validity of this 'strange' observation we compared the results obtained with the synthetic ground motions with corresponding results obtained using natural ground motions. This comparison is shown in Figure 10, where the $u_{top}$ displacements for 30 ground motions from the NGA PEER database [27], recorded in distances ranging from 17 to 23 km are plotted. For every natural earthquake considered, the results for both horizontal components are shown, resulting to 60 records in total. In the same plot, the line that corresponds to the mean values of $u_{top}$ for $R$=20 km, that was obtained for the synthetic records, is also shown. It is evident that the same behaviour is also observed for the natural ground motions: the maximum $u_{top}$ demand occurs for $M_w$=6.5, while for higher magnitudes the demand gradually decreases as in the case of the synthetic records. It is interesting to note that most of the points that correspond to natural earthquakes lie below the line of the synthetic ground motions. This was expected, since the synthetic records were constructed considering the directivity pulse with its maximum amplitude, i.e. typically that of the fault-normal direction; however natural ground motions were, in general, recorded in various directions with respect to the fault, and thus contain directivity pulses of reduced amplitude.

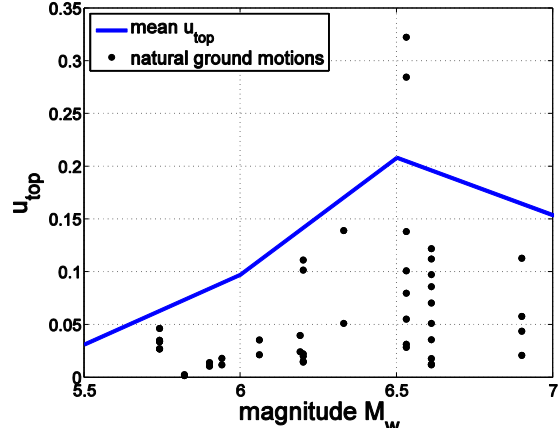

Figure 10. Comparison of the mean $u_{top}$ displacements produced using the adopted synthetic ground motion model for distance $R$ equal to 20 km against the corresponding results using natural ground motions.

For the same reason, the two components of each earthquake produced different values of $u_{top}$, with the larger value corresponding to the component that is close to the fault-normal direction and the lower one to the component that is close to the fault-parallel direction in which the directivity pulse is less strong.

The decrease in the amplitude of the response for ground motions that correspond to earthquakes of large magnitude is also evident in Fig. 11a, in which the dependence of $u_{top}$ on the pulse period, $T_p$, is depicted for all non-collapsed cases. As obvious from the envelope of the response (dashed line), initially, $u_{top}$ increases with the period $T_p$ as expected. This trend, however, reverses for periods $T_p$ longer than about 3 sec when $u_{top}$ generally decreases as $T_p$ increases. It is interesting to notice that the values of $u_{top}$ that correspond to pulse periods larger than 9 sec (i.e. produced by earthquakes of very large magnitude) are quite small, less than 0.35, classifying thus the performance to the first level of *damage limitation* (Table III).

Another interesting consequence of this phenomenon, which is caused by the saturation of *PGV*, is shown in Figs. 11b&c. In this case, the results are shown in terms of *PGA* versus $f_p$ and $A_p$ versus $f_p$ respectively, where *PGA* is peak acceleration of the combined record (low- and high-frequency components) and $A_p$ is the acceleration amplitude of the pulse alone, while $f_p = 1/T_p$. Results for $M_w < 6.5$ are not shown, because the column does not collapse for such earthquakes. Drawing the lower borderline (dashed line) between the non-collapsed (open circles) and collapsed (crosses) cases in Fig. 11b, the threshold that separates the safe and the unsafe areas can be defined. The fact that there are combinations of $PGA - f_p$ above this line that do not cause collapse of the column was expected, since it is known that increasing the amplitude of an earthquake that causes collapse to a column does not necessarily produce collapse, too (Psycharis *et al.* [5]). In this sense, the dashed line in Fig. 11b corresponds to the lower limit between the safe and unsafe areas.

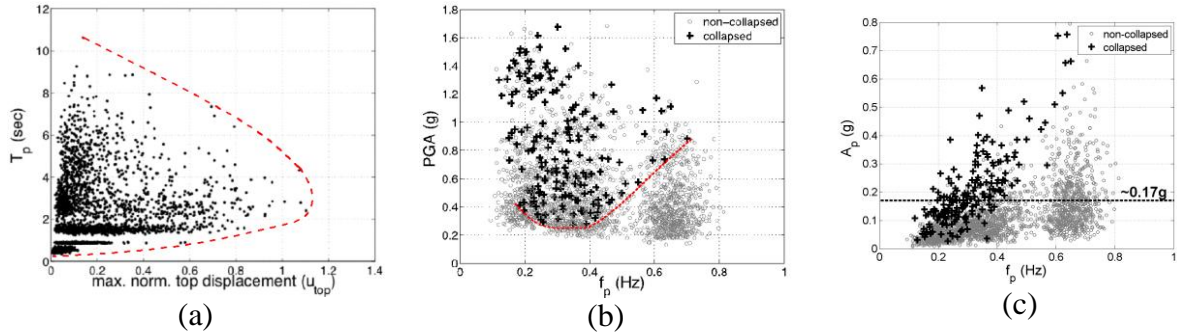

Figure 11. (a) Dependance of the maximum normalised top displacements, $u_{top}$, on the period of the pulse, $T_p$; (b) & (c) threshold between safe (non-collapse) and unsafe (collapse) regions on the $PGA$–$f_p$ plane and the $A_p$–$f_p$, respectively. The horizontal dashed line in (c) indicates the minimum base acceleration required to initiate rocking at the base of the column.

It is seen from Fig. 11b that for 0.25 H $<f_p<$ 0.4 Hz the safe − unsafe threshold is practically constant, while the required *PGA* to overturn the column increases almost linearly with $f_p$ for $f_p>$ 0.4 Hz. In the latter case, a similar increase in the required value of $A_p$ is observed in Fig. 11c. This was expected from previous analyses of the seismic response of classical columns [5], and also from the investigation of the toppling of rigid blocks to pulse-type excitations [28,29]. What is interesting, however, is that a similar increase in the required *PGA* to cause collapse of the column with the decrease of the pulse frequency to values less than 0.25 Hz (i.e. for $T_p>$ 4.00 sec) is observed in Fig. 11b. In this range of pulse periods, the corresponding acceleration amplitudes of the pulses are quite low (Fig. 11c), due to the saturation of the pulse velocity mentioned above (see Fig. 9b). It seems, therefore, that, for earthquakes of large magnitude, the peak acceleration of the higher-frequency component is crucial to the collapse or not of the column, something that was not realized up to now. It is reminded that, in such cases, a minimum value of *PGA* is required to initiate rocking of the drums, as the peak acceleration of the pulse is not strong enough. This is shown in Fig. 11c, in which the required *PGA* to initiate rocking at the base of the column (equal to 0.17 g) is shown with a dashed line. This corresponds to the minimum value required for initiation of rocking at any joint if the column behaves as a rigid block before rocking. It is noted however that, since the column is not rigid but flexible, ground motions with *PGA* smaller than 0.17 g can also trigger rocking at upper joints, depending on the ground motion characteristics. In any case, the threshold is not expected to be much lower than the dashed line in Fig. 11c, which means that pulses of $f_p<$ 0.4 Hz with $A_p$ much smaller than the threshold for rocking initiation are capable to cause collapse, provided that the column has already been set to rocking mode due to the higher-frequency component of the base motion.

Figure 12 shows the fragility surfaces of the classical column for the three performance levels of Table III ranging from *damage limitation* ($u_{top}>$ 0.15) to *significant damage* ($u_{top}>$ 1). It is reminded that $u_{top}>$ 0.15 means that the maximum top displacement during the ground shaking is larger than 15% of the base diameter and $u_{top}>$ 1 corresponds to intense rocking, close to collapse or actual collapse. When *damage limitation* is examined, the exceedance probability is of the order of 0.2 for $M_w$ = 6 and increases rapidly for ground shakings of larger magnitude. For the worst scenario among those examined ($M_w$ = 7.5, $R$ = 5 km), the probability that the top displacement is larger than 15% of $D_{base}$ is equal to unity, while in the range $M_w$ = 6.5-7.5 and $R >$ 15 km a decrease in the exceedance probability is observed as discussed in the previous paragraphs. Similar observations hold for the exceedance of the *significant damage* limit state ($u_{top}>$ 0.35), but the probability values are smaller. For the *near collapse* limit state ($u_{top}>$ 1.0), the probability of exceedance reduces significantly for large distances, even for large magnitudes. It is interesting to note that the $u_{top}>$ 1.0 surface practically coincides with the probability of collapse of Figure 9, which shows that, if the top displacement reaches a value equal to the base diameter, there is a big possibility that the column will collapse a little later.

Figure 13 shows the fragility surfaces when the *EDP* is the normalized permanent drum dislocation, $u_d$, and considering the performance levels of Table IV. For the *limited deformation* limit state ($u_d > 0.005$), probabilities around 0.3 are observed for magnitudes close to 6. Note that, for the column of the Parthenon with an average drum diameter about 1600 mm (Figure 2), $u_d > 0.005$ refers to residual displacements at the joints exceeding 8 mm. The probability of exceedance of the *light deformation* performance criterion ($u_d > 0.01$), which corresponds to residual drum dislocations larger than 16 mm, is less than 0.2 for all earthquake magnitudes examined and for distances from the fault larger than 10 km. The *significant deformation* limit state ($u_d > 0.02$) was exceeded only in a few cases.

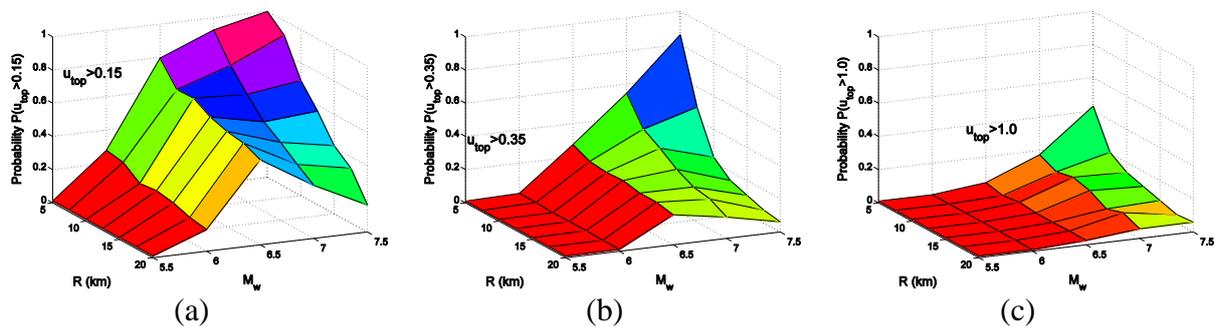

Figure 12. Fragility surfaces with respect to the maximum capital displacement $u_{top}$ for the performance levels of Table III: (a) $u_{top} > 0.15$; (b) $u_{top} > 0.35$; (c) $u_{top} > 1.0$.

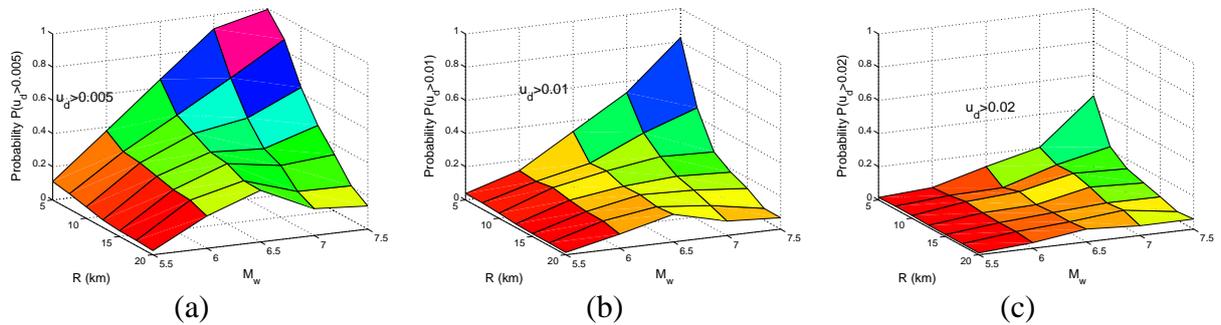

Figure 13. Fragility surfaces with respect to the permanent drum dislocations, $u_d$ for the performance levels of Table 3: (a) $u_d > 0.005$; (b) $u_d > 0.01$; (c) $u_d > 0.02$.

In Figs 14 and 15 the *PGA* and *PGV* are plotted versus the *EDP*s considered, $u_{top}$ and $u_d$. The scatter in the results is significant in both cases, slightly smaller for *PGV*. However, clear trends can be identified in the response, especially from Figure 15, showing in average a generally linear relation between the deformation (maximum and residual) with *PGV*.

Another conclusion is that very strong earthquakes, with *PGV* that exceeds 150 cm/sec, are required for bringing the column of the Parthenon close to collapse ($u_{top} > 1$). However, significant dislocations of the drums ($u_d > 0.02$) can occur for weaker earthquakes with *PGV* > 40 cm/sec. These observations are in accordance with findings of previous studies (Psycharis *et al*. [9]).

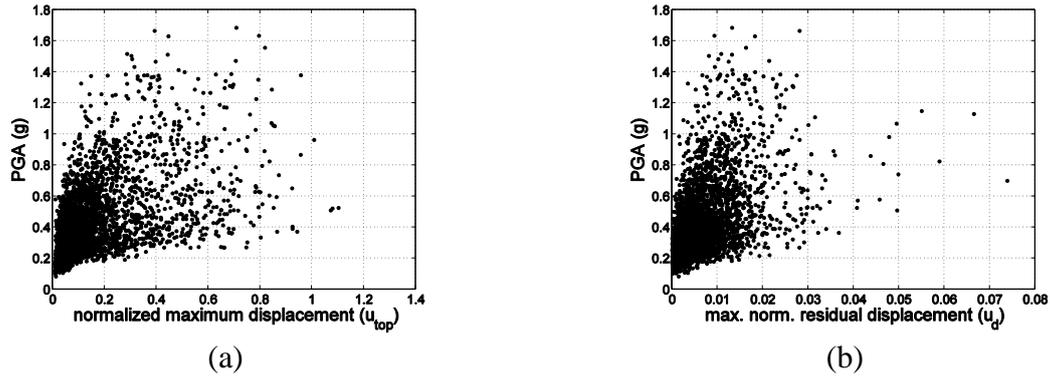

(a)                                                (b)

Figure 14. Scatter plots of *PGA* versus: (a) maximum normalized displacement $u_{top}$; (b) maximum normalized displacement $u_d$.

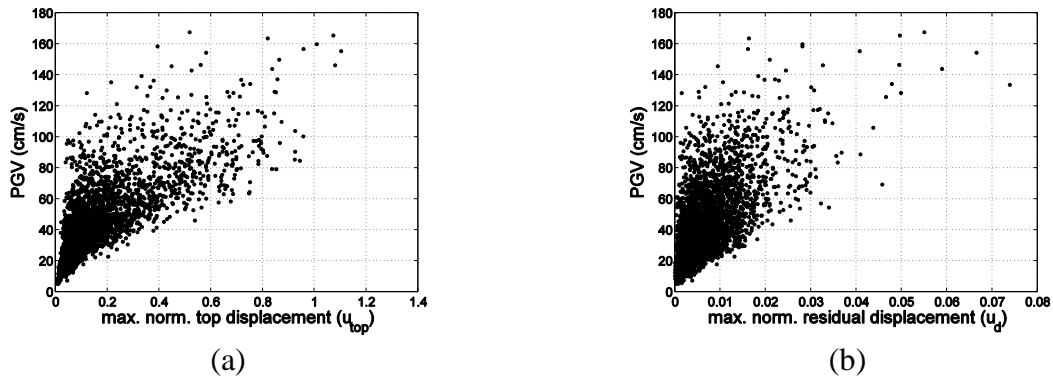

(a)                                                (b)

Figure 15. Scatter plots of *PGV* versus: (a) maximum normalized displacement $u_{top}$; (b) maximum normalized displacement $u_d$.

Finally, fragility curves for the *EDP*s thresholds defined in Tables III and IV and using *PGA* and *PGV* as intensity measures, are shown in Figure 16. The curves were obtained using the procedure that is schematically shown in Figure 3 and assuming that the y-axis is divided to 12 stripes of equal width. It is seen that the probability that a moderate earthquake with *PGA* ~ 0.3 g and *PGV* ~ 40-50 cm/sec has only 10% probability to cause considerable rocking to the column with $u_{top}$> 0.35 and to produce permanent dislocations of the drums that exceed 1% of their diameter. Unfortunately, the existing drum dislocations of the columns of the Parthenon, which are of this order of magnitude, cannot be directly compared with the $u_d$ values of the analyses, not only because the columns are not free-standing as assumed in this investigation, but also because it is not certain that the existing drum displacements were caused by earthquakes solely or by the big explosion that occurred in 1687 and shook considerably the whole structure. Additionally, the drum dislocations that are measured today in the monument are probably the cumulative effect of a number of earthquakes rather than the result of a single strong event.

## CONCLUSIONS

A seismic risk assessment of a column of the Parthenon Pronaos is performed using Monte Carlo simulation with synthetic ground motions which contain a high- and a low-frequency component. The ground motions considered combine the stochastic method and a simple analytical pulse model to simulate the directivity pulse contained in near source records. The response of the column was calculated for 35 $M_w$–$R$ scenarios with magnitudes $M_w$ ranging from 5.5 to 7.5 and distances $R$ from the fault in the range of 5 to 20 km. For every $M_w$–$R$ scenario 100 Monte Carlo Simulations (MCS) were performed resulting to 3500 simulations in total.

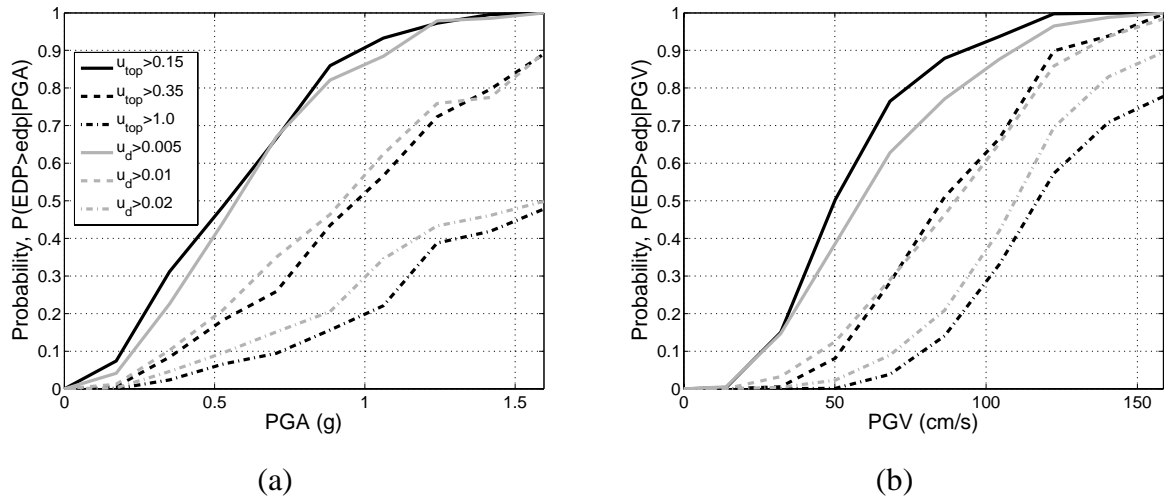

(a)              (b)

Figure 16. Fragility curves using different intensity measures: (a) peak ground acceleration; (b) peak ground velocity.

Two engineering demand parameters (*EDP*s) are adopted for the assessment of the vulnerability of classical columns: (a) the maximum displacement at the capital normalized by the base diameter; and (b) the relative residual dislocation of adjacent drums normalized by the diameter of the corresponding drums at their interface. Three performance levels are assigned to each *EDP* and the values of the corresponding thresholds are proposed.

The fragility analysis demonstrated some of the salient features of these spinal systems under near-fault earthquake excitations which were not realized up to now. The conclusions can be summarized as follows.

- In general, strong ground motions lead to large top displacements during the strong shaking and also produce large permanent deformation of the column. However, there is significant scattering of the results indicating that intense rocking does not necessarily imply large residual dislocations of the drums and also that large drum dislocations can occur for relatively weak shaking of the column.
- For small distances from the fault, less than 10 km, the mean values of $u_{top}$ and $u_d$ increase monotonically with the magnitude as expected. However, for larger distances from the fault, the maximum values of these parameters occur for magnitude $M_w \sim 6.5$, while for larger magnitudes the top displacement and the residual deformation decrease. This counter-intuitive response, which was verified for real earthquake records, is attributed to the saturation of the *PGV* for magnitudes larger than 7 and the resulting small acceleration amplitude of the directivity pulses, which is not capable to produce intense rocking. Due to this phenomenon, the up-to-now belief that the vulnerability of classical monuments increases with the predominant period of the excitation does not hold for very long periods (longer than 4.0 sec for the column of the Parthenon), for which the response of the columns generally decreases as the period $T_p$ increases.
- In addition to the above conclusion, and for earthquakes of large magnitude containing pulses of long period (larger than 4.0 sec for the case of the Parthenon column), the *PGA* of the high-frequency component of the ground motion seems to be crucial to the collapse of the column, with larger values of *PGA* needed as $f_p$ decreases.
- The fragility surfaces produced for the adopted *EDP*s and for all performance levels showed that very strong ground motions are required to bring the column close to collapse and cause significant drum dislocations. On the contrary, moderate earthquakes with *PGA* ~ 0.3 g and *PGV* ~ 40-50 cm/sec have only 10% probability to cause considerable rocking to the column ($u_{top}$ > 0.35) and to produce permanent dislocations of the drums that exceed 1% of their diameter. This was expected from previous analyses of classical columns under earthquake excitations.

- Significant scatter in the results was observed when the intensity measures *PGA* and *PGV* were plotted versus the *EDP*s $u_{top}$ and $u_d$, with *PGV* being a little better *IM* than *PGA*. However, clear trends can be identified in the response, showing a generally linear relation between the deformation (maximum and residual) with *PGV*.

## ACKNOWLEDGEMENTS

Partial financial support for this study has been provided by the EU research project "DARE" ("Soil-Foundation-Structure Systems Beyond Conventional Seismic Failure Thresholds: Application to New or Existing Structures and Monuments"), which is funded through the 7$^{th}$ Framework Programme "Ideas", Support for Frontier Research – Advanced Grant, under contract number ERC-2—9-AdG 228254-DARE to professor G. Gazetas.